\documentclass[journal]{IEEEtran}
\ifCLASSINFOpdf
\else
\fi
\usepackage{amsmath}
\usepackage{array}

\usepackage{stfloats}
\ifCLASSOPTIONcaptionsoff
  \usepackage[nomarkers]{endfloat}
 \let\MYoriglatexcaption\caption
 \renewcommand{\caption}[2][\relax]{\MYoriglatexcaption[#2]{#2}}
\fi
\usepackage{url}

\usepackage{color}
\usepackage{soul}
\usepackage{graphicx}
\usepackage{amssymb}
\usepackage{pifont}
\usepackage{subcaption}
\usepackage{multirow}
\usepackage{algorithm}
\usepackage[noend]{algpseudocode}
\usepackage{comment}
\usepackage{soul}

\begin{document}
%
\title{RoadNet-RT: High Throughput CNN Architecture and SoC Design for Real-Time Road Segmentation}
%
%
%

\author{Lin~Bai,~\IEEEmembership{Student Member,~IEEE,}
        Yecheng~Lyu,~\IEEEmembership{Student Member,~IEEE,}
        and~Xinming~Huang,~\IEEEmembership{Senior Member,~IEEE}
\thanks{This work is partially supported by U.S. NSF Grant CCF-2006738 and by The MathWorks fellowship. The authors are with the Department of Electrical and Computer Engineering, Worcester Polytechnic Institute, Worcester, MA 01609, USA. The corresponding author is Xinming Huang (e-mail: xhuang@wpi.edu).}
\thanks{Manuscript received }}

%
%

\markboth{Journal of \LaTeX\ Class Files,~Vol.~14, No.~8, August~2015}%
{Shell \MakeLowercase{\textit{et al.}}: Bare Demo of IEEEtran.cls for IEEE Journals}
%



\maketitle

\begin{abstract}
In recent years, convolutional neural network (CNN) has gained popularity in many engineering applications especially for computer vision. In order to achieve better performance, more complex structures and advanced operations are incorporated into neural networks, which results in very long inference time. For time-critical tasks such as autonomous driving and virtual reality, real-time processing is fundamental. In order to reach real-time processing speed, a lightweight, high-throughput CNN architecture namely RoadNet-RT is proposed for road segmentation in this paper. \textcolor{black}{It achieves 92.55$\%$ MaxF score on KITTI road segmentation dataset. The inference time is about 9 ms per frame when running on GTX 1080 GPU. Comparing to the state-of-the-art network, RoadNet-RT speeds up the inference time by a factor of 17.8 at the cost of only \textcolor{black}{3.75$\%$} loss in accuracy.} What is more, on CamVid dataset its accuracy is 92.98$\%$. \textcolor{black}{Several techniques such as depthwise separable convolution and non-uniformed kernel size convolution are optimized in the hardware accelerator design. The proposed CNN architecture has been successfully implemented on a ZCU102 MPSoC FPGA that achieves the computation capability of 331 GOPS using INT8 quantization. The system throughput reaches 196.7 frames per second with input image size of 280$\times$960. The source code is published at \url{https://github.com/linbaiwpi/RoadNet-RT.}}
\end{abstract}


%
\IEEEpeerreviewmaketitle

\section{Introduction}
\IEEEPARstart{N}{owadays} autonomous vehicles have become one of the most promising technologies. Owing to the continuous development of Convolutional Neural Networks (CNNs), many recent research were focused on improving the accuracy performance of the perception system for autonomous vehicles, such as vehicles or pedestrians detection \cite{du2018general}\cite{yang2019std}, depth completion \cite{cheng2019cspn++}, road segmentation \cite{sun2019reverse}\cite{chen2017rbnet} and object tracking \cite{choi2015near}. However, most of these neural networks are very deep with a huge number of parameters. Even running on a state-of-the-art GPU, few of them are able to process sensor data in real-time. This prevents them being applied to time-critical tasks such as autonomous driving. Therefore, a fast lightweight CNN with reasonable accuracy is valuable to those time-critical applications.

\begin{figure}[t]
	\centering
	\includegraphics[width=0.99\columnwidth]{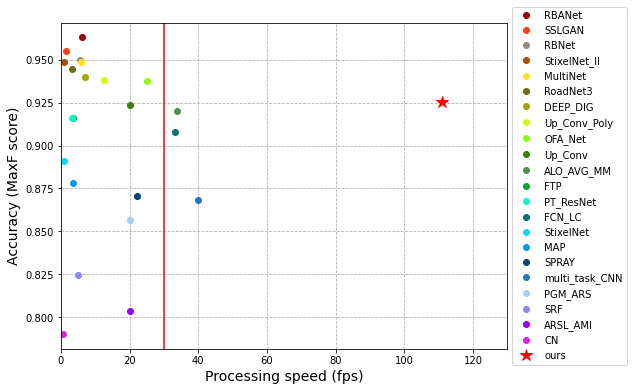}
	\caption{Processing speed v.s. accuracy on the KITTI road segmentation test dataset. Red star indicates our method, and colored dots represent other methods. All of these solutions are tested on GPU/CPU which listed in KITTI leaderboard of road segmentation task. Red line is the border of real-time.}
	\label{fig:performance}
\end{figure}

Road segmentation is one of the fundamental perception tasks for autonomous driving, which tells the vehicles where the drivable region is. This task has been well studied by many researchers concerning to the accuracy performance measured by benchmarks. While as a time-critical task, only 3 of the existed methods are able to process in real-time as illustrated in Fig.~\ref{fig:performance}, where the red line indicates the real-time processing speed at 30 frames per second (fps), and none of their throughput exceeds 40 fps. As a fundamental task prior to path planning and dynamic control, road segmentation is expected to process input images at a much faster frame rate, such that it guarantees the real-time response of an autonomous driving system. Thus, there is an urgent need of real-time road segmentation that can process each image within a very short time while maintaining good accuracy, which bridges the gap between academic research and industry practice.


In this paper, we propose RoadNet-RT, a real-time road segmentation network, which is able to run in real-time on a GPU. Besides, we have summarized some optimization techniques aiming to convert ordinary CNN structures into hardware friendly ones. As a demonstration, RoadNet-RT has been successfully implemented on an FPGA by applying these techniques, resulting real-time processing on hardware.
The contributions of this paper are summarized as following:
\begin{itemize}
	\item \textcolor{black}{A lightweight high throughput CNN named RoadNet-RT is proposed, whose segmentation accuracy is 92.55$\%$ on KITTI road segmentation leaderboard. RoadNet-RT extracts features from two branches, one shallow branch for spatial information and one deep branch for context information. Its inference time on NVIDIA GTX 1080 is about 9 ms. When comparing to the state-of-the-art RBANet \cite{sun2019reverse}, this network reduces the inference time by 94.4\%, with only \textcolor{black}{3.75$\%$} loss in accuracy.}
	\item Aimed at providing the general guidelines on how to transform a segmentation CNN into a hardware friendly one with both computation and bandwidth efficiencies, we investigate several hardware optimization techniques through a series of experiments with quantitative results. For instance, how to employ depthwise separable convolution, how to deal with convolutions with different kernel size and dilated convolution, and whether using batch normalization are studied.
	\item \textcolor{black}{An efficient hardware accelerator has been implemented on a ZCU102 MPSoC FPGA platform. By balancing the bandwidth and computation capability, this accelerator can process 196.7 image frames per second with INT8 quantization, equivalent to the efficiency of 331 Giga Operations Per Second (GOPS).} 
\end{itemize}

The rest of the paper is organized as following: Sec.~II summarizes the existing research on road segmentation, real-time segmentation CNNs and the FPGA implementations of segmentation networks. In Sec.~III, the proposed segmentation network model is described together with its training details. An in-depth study of network optimization techniques for hardware efficiency and accuracy performance is presented in Sec.~IV. The FPGA implementation and its results are discussed in Section V and VI, respectively. Sec.~VII concludes the entire paper.


\section{Related Work}

\subsection{Road segmentation}
Lots of research efforts have been paid on road segmentation task in KITTI. The RBANet proposed in \cite{sun2019reverse} adopted the classical encoder-decoder structure. Instead of using the direct skip connection in U-Net \cite{ronneberger2015u} and SegNet \cite{badrinarayanan2017segnet}, a residual refinement module bridged encoder and decoder parts, which consisted of reversed attention and boundary attention mechanisms. So that high-resolution spatial details were preserved for decoding. Atrous Spatial Pyramid Pooling (ASPP) module was also utilized in RBANet. For images size $360\times 720$ running on GTX Titan XP, the processing time was 0.16 second per frame. In \cite{han2018semisupervised}, 
SSLGAN served to train unlabeled data and enhanced road feature representations using a discriminator from GAN. Labeled data contain many redundant areas, so training both labeled and unlabeled data prevents the overfitting problem and accelerates the convergence speed. Its processing speed was 0.7s per frame on TITAN X. A road and road boundary detection network (RBNet) was proposed in \cite{chen2017rbnet}. Based on a Bayesian network, RBNet could simultaneously estimate the probabilities of a pixel on the image belonging to the road and road boundary so that the road and road boundary detection were combined into a single process. It was able to process each frame in 0.18s on Tesla K20c (5 GB). StixelNet \cite{garnett2017real} posed generic static obstacles represented as stixels and learnt directly using a CNN. StixelNet II \cite{garnett2017real} was a unified network with real-time detection capability for both categorized and un-categorized objects. This network performed well on column-based obstacle detection and road segmentation but was not sensitive to the distinction of road boundaries. MultiNet \cite{teichmann2018multinet} utilized the same encoder which was based on VGG16 to supply features to different decoders for classification, segmentation, and detection tasks. In segmentation decoder, the low-resolution segmentation feature map was convoluted and then upsampled using transposed convolution. It was claimed that MultiNet could perform inference at 23 fps. The structure of Up-Conv-Poly \cite{leivas2016ef} was very similar to U-Net. It achieved MaxF score 93.83$\%$. For images with size 500$\times$500, this network could process each frame within 83 ms on TITAN X GPU.

Other CNN based road segmentation algorithms such as DEEP-DIG \cite{munoz2017deep} and MAP \cite{laddha2016map} generated a precise drivable region but required heavy computational power.

In our previous work RoadNetV3 \cite{lyu2019road}, we introduced Long-Short Term Memory (LSTM) to help finding the contour of the road. It extracted features via a FCN-like encoder. After that, several convolutional-LSTM layers followed to predict the contours of drivable region. It achieved 93.08$\%$ in accuracy but 300 ms per frame.

\subsection{Real-time segmentation}
\begin{figure}[htbp]
    \centering
    \begin{subfigure}[b]{0.35\columnwidth}
            \includegraphics[width=\columnwidth]{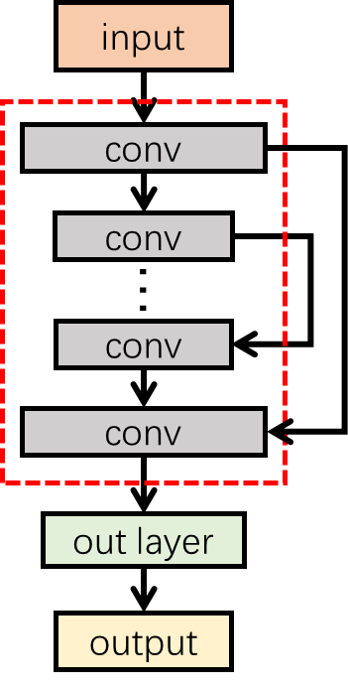}
            \caption{}
    \end{subfigure}
    \begin{subfigure}[b]{0.55\columnwidth}
            \includegraphics[width=\columnwidth]{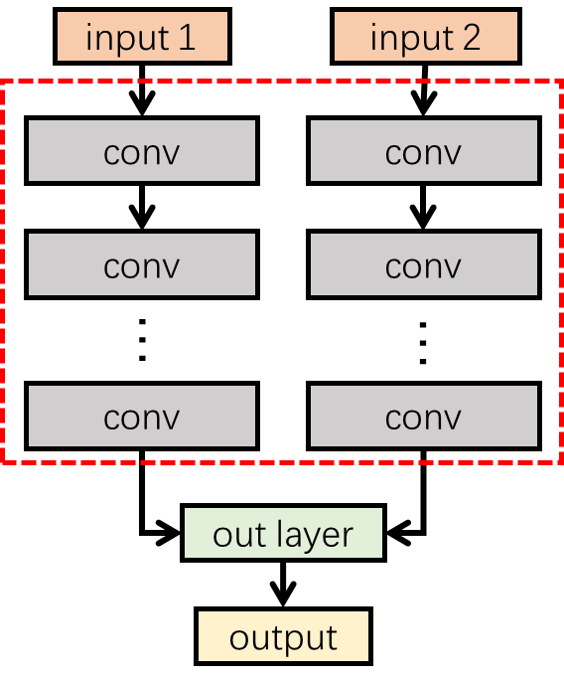}
            \caption{}
    \end{subfigure}
    \caption{The mainstream structures for real-time semantic segmentation. (a) illustrates the u-shape encoder-decoder structure and (b) demonstrates the bilateral structure}
    \label{fig:seg_struct}
\end{figure}

In recent years, some researchers have shifted their focus to real-time segmentation tasks. Their solutions are generally categorized into two groups (Fig.~\ref{fig:seg_struct}), one is encoder-decoder network and the another one is bilateral network.

FPENet \cite{liu2019feature} adopted the encoder-decoder structure. By using a feature pyramid encoding block to encode multi-scale contextual features with depthwise dilated convolutions in all stages and a mutual embedding upsample module as decoder, FPENet efficiently aggregated of high-level semantic features and low-level spatial details. Through introducing an efficient spatial pyramid (ESP), ESPNet \cite{mehta2018espnet} brought great improvement in both speed and performance. In its improved version, ESPNet-V2 \cite{mehta2019espnetv2} further enlarged the receptive field and reduced the calculation of parameters. In \cite{lidabnet}, DABNet balanced the efficiency and accuracy via stacking lightweight blocks with different dilation rates. DFANet \cite{li2019dfanet} aggregated multi-scale features from different layers to gain higher accuracy in spatial details. The lightweight backbone of DFANet guaranteed its real-time processing speed.

ContextNet \cite{PoudelBLZ18} proposed the solution of bilateral structure for the first time. A deep but low-resolution network extracted the context information. And a shallow but high-resolution network focused on detailed spatial information. BiSeNet \cite{yu2018bisenet} inherited the solution of ContextNet and improved the feature fusion modules by creating attention residual module and feature fusion module. Via adding global pooling layer and residual layer, BiSeNet outperformed ContextNet. In ICNet \cite{zhao2018icnet}, the authors borrowed the image pyramid thinking from PSPNet \cite{zhao2017pyramid}. One more branch was added to acquire more spatial details. Plus, the label guided training for each branch, ICNet had better accuracy than BiSeNet but longer processing time. BiSeNet-V2 \cite{yu2020bisenet} improved the first version by replacing feature fusion module into aggregation module and using Seg Head to guide the loss of each feature extractor layer. Other networks like LBN-AA \cite{9042876}, CANet \cite{tang2020attention} also used similar structure.

Solutions other than the two mentioned above also represent good results. FarSee-Net \cite{zhang2020farsee} applied Cascaded Factorized Atrous Spatial Pyramid Pooling (CF-ASPP) at the end of feature extraction layers to guarantee enough spatial information was captured. What is more, to reduce the number of operations, sub-pixel convolution was deployed, so that FarSee-Net accepted low-resolution input and generated high-resolution output.

\subsection{FPGA implementation of segmentation}
To accelerate the inference speed, a great amount of effort focused on FPGA implementation of segmentation neural networks. The key to hardware accelerator for CNNs was the trade-off between bandwidth and computation capability. U-Net \cite{ronneberger2015u} and FCN \cite{long2015fully} are both implemented in \cite{liu2018optimizing}. By utilizing convolution plus board removing method, this accelerator operated transposed convolution efficiently. Its performance was 107 GOPS and supported up to 17 fps for 512$\times$512 images. A straight-forward fully convolution neural network for segmentation has been proposed and implemented on FPGA \cite{lyu2018real}\cite{lyu2018chipnet}. Without changing the channel depth for each layer and skip connections used in U-Net \cite{ronneberger2015u}, this accelerator pushed its performance to process 79.4 fps for input size 64$\times$180$\times$14. Liu merged the convolution and transposed convolution into one vector multiplication unit and fused all intermediate feature maps in on-chip memory \cite{liu2019towards}. And the FPGA implementation reached 1578 GOPS, which was 57 fps for 256$\times$256$\times$3 images. Another hardware architecture combining the convolution and transposed convolution operations was proposed in \cite{bai2020deconv}. Its computation capability was 151.5 GOPS and 94.3 GOPS for convolution and transposed convolution respectively. besides, a 3D segmentation CNN accelerator was implemented in \cite{shen2019scale}.

\section{Proposed Network}
\begin{figure}[htbp]
	\centering
	\includegraphics[width=0.9\columnwidth]{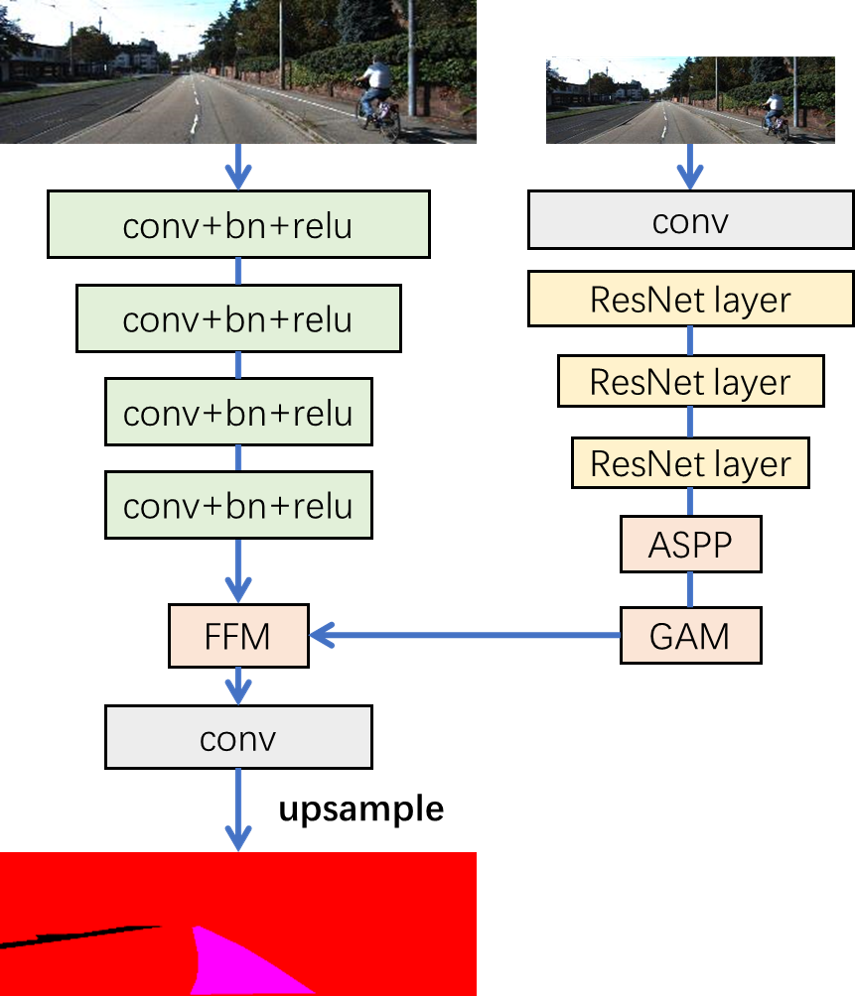}
	\caption{\textcolor{black}{Real-time road segmentation network RoadNet-RT structure}}
	\label{fig:roadnet}
\end{figure}

The proposed road segmentation network is inspired by ContextNet \cite{PoudelBLZ18}, BiSeNet \cite{yu2018bisenet} and ICNet \cite{zhao2018icnet}. It consists of two branches for context information and spatial information extraction respectively, as shown in Fig.~\ref{fig:roadnet}.

\begin{figure}[htbp]
	\centering
	\includegraphics[width=0.8\columnwidth]{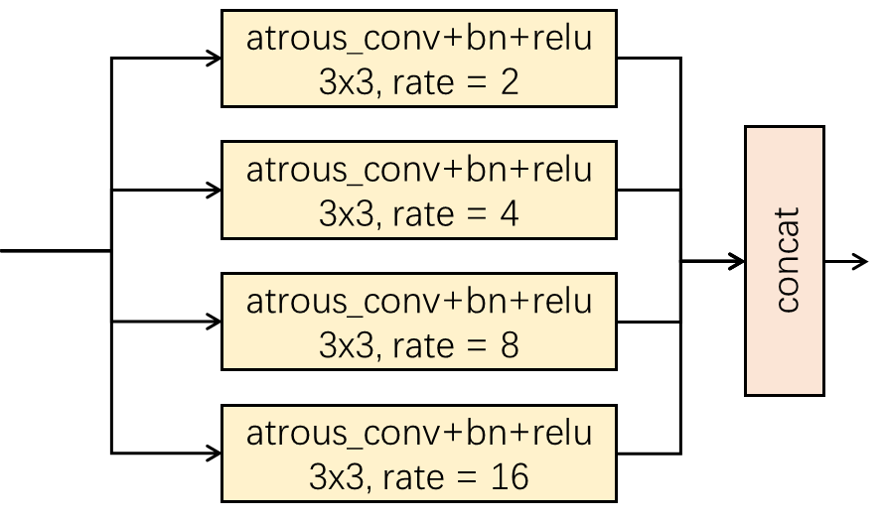}
	\caption{\textcolor{black}{Structure of ASPP}}
	\label{fig:aspp}
\end{figure}

\begin{figure}[htbp]
    \centering
    \begin{subfigure}[b]{0.35\columnwidth}
            \includegraphics[width=\columnwidth]{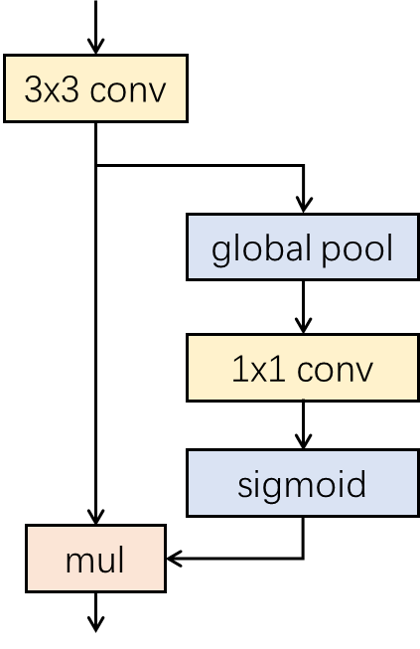}
            \caption{}
            \label{fig:arm}
    \end{subfigure}
    \begin{subfigure}[b]{0.4\columnwidth}
            \includegraphics[width=\columnwidth]{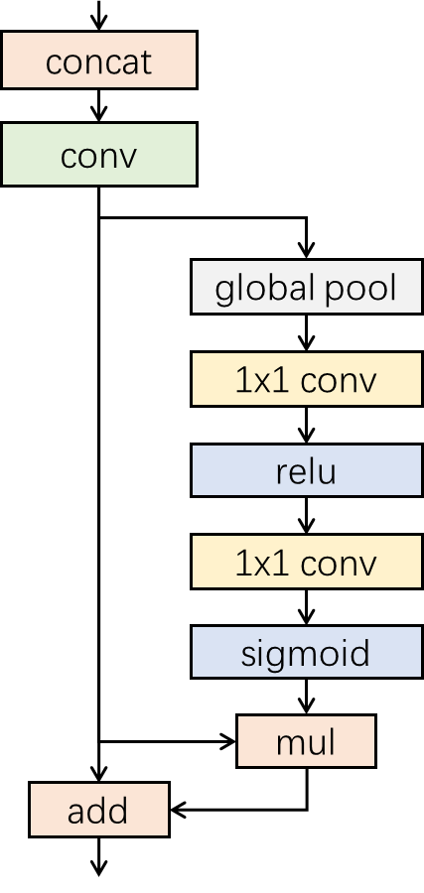}
            \caption{}
            \label{fig:ffm}
    \end{subfigure}
    \caption{\textcolor{black}{(a) structure of GAM, (b) structure of FFM \cite{yu2018bisenet}}}
    \label{fig:ffm_arm}
\end{figure}

\textcolor{black}{The context branch is a deep network for extracting the context information, which consists of an input convolutional layer and two residual modules from ResNet18 \cite{he2016deep}. Subsequently, the extracted features are fed to the ASPP module in order to concatenate the features from different fields of perception (dilated rates are 2, 4, 8 and 16, depth are 32 for each, in \textcolor{black}{Fig.~\ref{fig:aspp}}). Next, a Global Attention Module (GAM) is introduced to refine the context information. The GAM (Fig.~\ref{fig:arm}) is modified from the Attention Refinement Module in \cite{yu2018bisenet}. The GAM consists of a global average pooling layer together with a 1$\times$1 convolutional layer who extracts global context feature. These refined global features are applied to context features via multiplication. The sigmoid layer decides whether to apply the global features or not. Since the context path does not have to focus on spatial details, we shrink the input image size by half in both width and height, as a step to further reduce the computation.}

For spatial path, which focuses on spatial details of the input images, contains only four convolution layers. To enhance its capability of noticing details, no image resize is applied here. \textcolor{black}{The context and spatial branches are fused in a residual refinement way, called Feature Fusion Module (FFM) \cite{yu2018bisenet} (Fig.~\ref{fig:ffm})}. The residual of FFM is the product of input feature map and its global attention path, including global average pooling layer, 1$\times$1 convolutional layer, activation layers (ReLU and Sigmoid). At the end of the network, to reproduce the output with the same size as input, the output of FFM is upsampled 8 times by the bi-linear resize algorithm.

The number of channels is chosen to be factor of 32. This is based on the number of parallelisms the hardware accelerator could support, in order to maximize the efficiency of it.

\subsection{Training Details}
This road segmentation network is implemented using Keras and trained from scratch on a single GeForce GTX 1080 GPU. All the convolutional layers were initialized using the Xavier uniform initializer \cite{glorot2010understanding}. During training, the batch size is set to 24. The Adam optimizer works with learning rate 1e-3. When in plateau, a reduction rate of 0.8 is applied to the learning rate. A hybrid loss function combining Dice loss and Focal loss is deployed here expecting to balance the positive and negative samples.

Data augmentation for training includes random horizontal flip, Gaussian noise adding, random brightness contrast, random blurring, etc.

\begin{figure*}[b]
    \centering
    \begin{subfigure}{0.5\textwidth}
            \includegraphics[width=\linewidth]{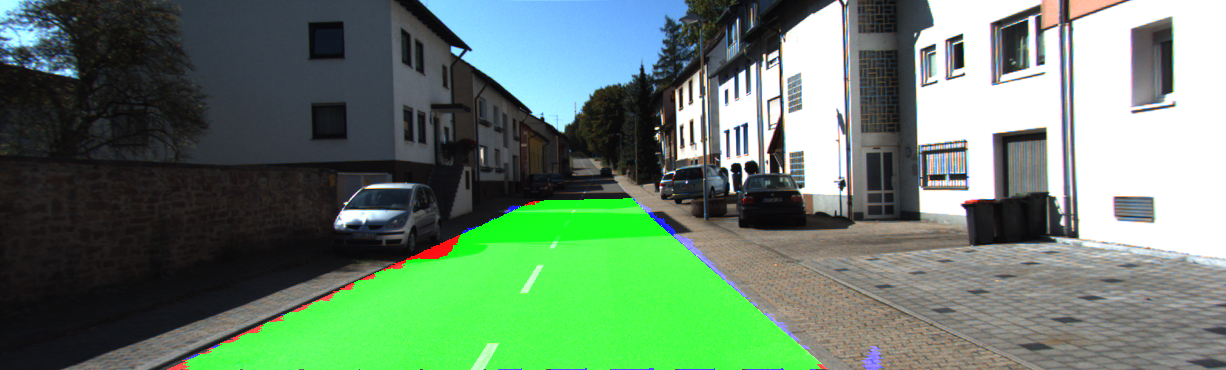}
    \end{subfigure}%
    \begin{subfigure}{0.5\textwidth}
            \includegraphics[width=\linewidth]{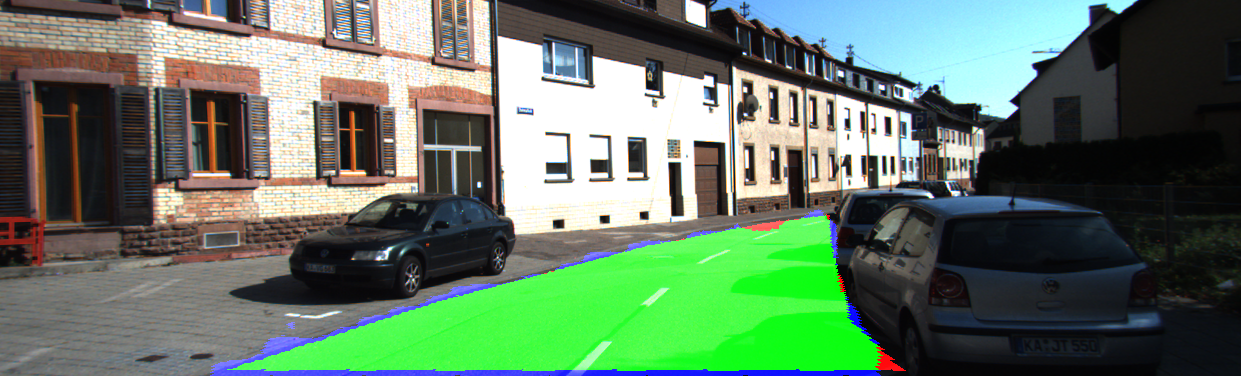}
    \end{subfigure}%
    \newline
    \begin{subfigure}{0.5\textwidth}
            \includegraphics[width=\linewidth]{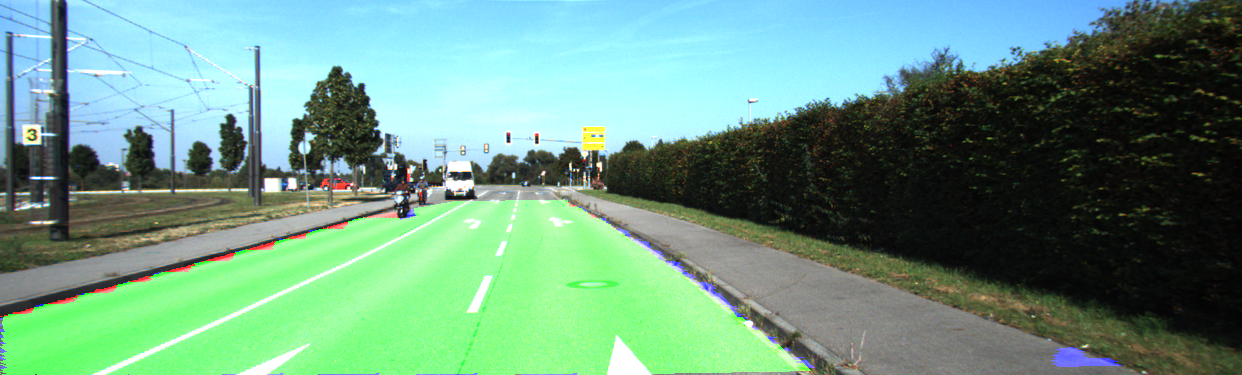}
    \end{subfigure}%
    \begin{subfigure}{0.5\textwidth}
            \includegraphics[width=\linewidth]{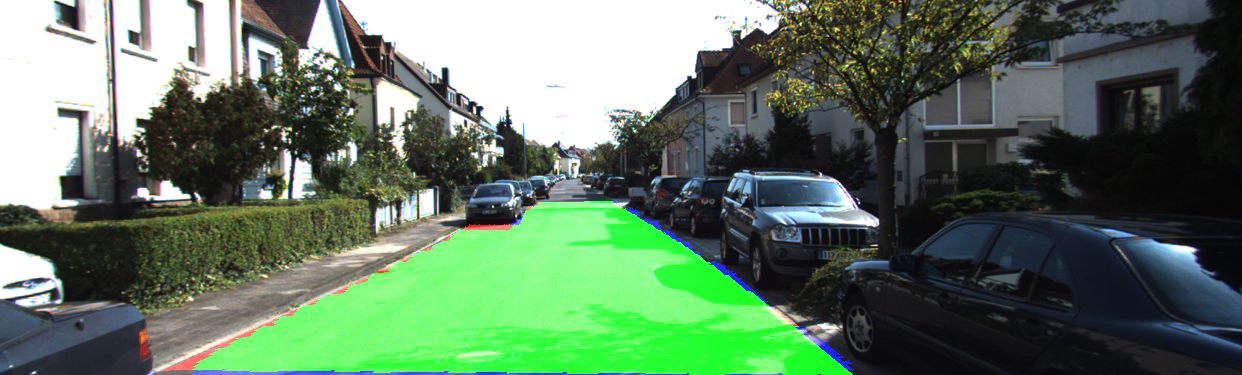}
    \end{subfigure}%
    \caption{Road segmentation results in camera view}
    \label{fig:pred}
\end{figure*}

\begin{figure*}[htbp]
    \centering
    \begin{subfigure}{0.25\textwidth}
            \includegraphics[width=\linewidth]{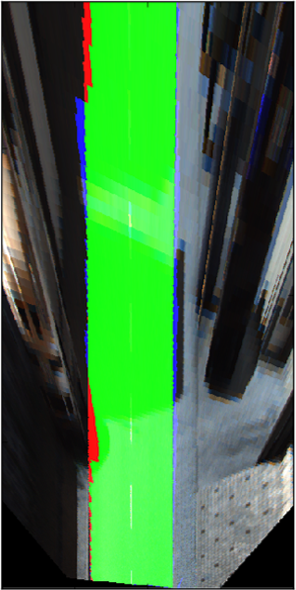}
    \end{subfigure}%
    \begin{subfigure}{0.25\textwidth}
            \includegraphics[width=\linewidth]{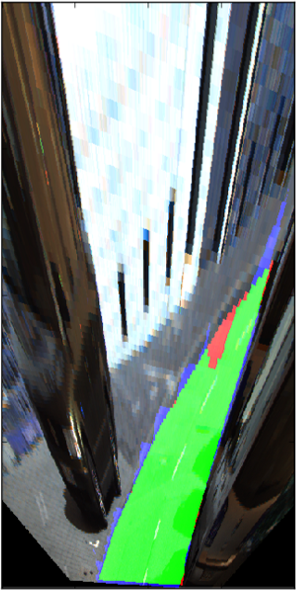}
    \end{subfigure}%
    \begin{subfigure}{0.25\textwidth}
            \includegraphics[width=\linewidth]{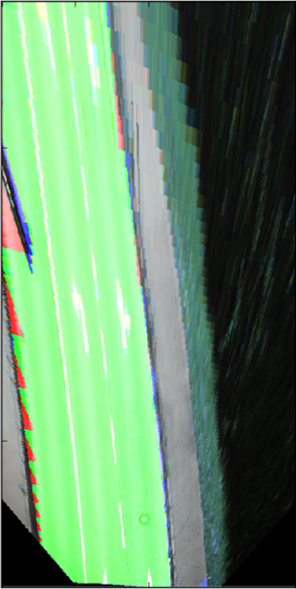}
    \end{subfigure}%
    \begin{subfigure}{0.25\textwidth}
            \includegraphics[width=\linewidth]{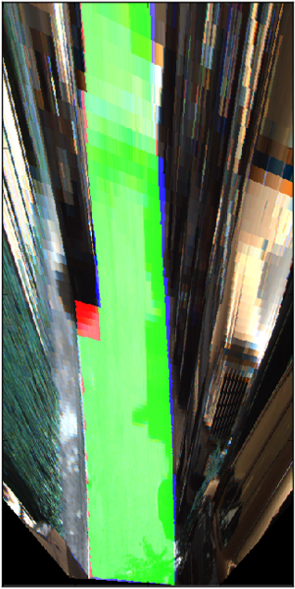}
    \end{subfigure}%
    \caption{Road segmentation results in Bird-Eye View}
    \label{fig:bev_pred}
\end{figure*}

\subsection{Dataset and Evaluation}
\textcolor{black}{\textbf{KITTI.}} The dataset for training and evaluation is the KITTI road segmentation dataset, which contains 289 training images and 290 testing images. The training image size ranges from 370$\times$1224 to 375$\times$1242. The evaluation job is done by an online evaluation server supplied by KITTI. The evaluation (Tab.\ref{tab:perform} is divided into Urban Unmarked (UU), Urban Marked (UM) and Urban Multiple Marked lanes (UMM). URBAN\_ROAD is the comprehensive evaluation of the above three.

When running on GeForce GTX 1080 GPU, this network can process each image with 280$\times$960 pixels in 9 ms. Four samples of predictions are demonstrated in front view and bird eye view by Fig.~\ref{fig:pred} and Fig.~\ref{fig:bev_pred} respectively, where green area represents the overlap between prediction and ground truth, red area is road in ground truth but not correctly predicted by our network, and blue area is not road but recognized as road by our network.

\textcolor{black}{Tab.~\ref{tab:kitti_cmp} shows the performance comparison among RoadNet-RT and other state-of-the-art networks. The FNR (False Negative Rate) reflects the ratio of pixels, which are road but are wrongly recognized as non-road. While the FPR (False Positive Rate) calculates the ratio of pixels, which are non-road but are wrongly classified as road. From Tab.\ref{tab:kitti_cmp}, we can see RoadNet-RT has much higher FNR (7.84\%) than the peers. Considering moving autonomous vehicles, high FNR would pose more restrictions on the drivable region. On the contrary, a high FPR means the neural network classifies more non-road pixels as road. For example, vehicles may recognize other cars on the roadside or bush as drivable region. Thus, high FPR would cause a safety issue. In FPR column of Tab.\ref{tab:kitti_cmp}, RoadNet-RT's FPR is comparable to the peers. Therefore, we consider Roadnet-RT is as safe as other state-of-the-art networks listed in Tab.~\ref{tab:kitti_cmp}.}

\begin{table}[htbp]
    \centering
    \caption{\textcolor{black}{KITTI evaluation comparison on URBAN\_ROAD benchmark}}
    \label{tab:kitti_cmp}
    \begin{tabular}{|c | c | c | c | c | c | c |}
    \hline
    \textbf{\textcolor{black}{Benchmark}} & \textbf{\textcolor{black}{MaxF}} & \textbf{\textcolor{black}{AP}} & \textbf{\textcolor{black}{FPR}} & \textbf{\textcolor{black}{FNR}}\\
    \hline
    \textcolor{black}{RBANet \cite{sun2019reverse}} & \textcolor{black}{96.30 \%} & \textcolor{black}{89.72 \%} & \textcolor{black}{2.75 \%} & \textcolor{black}{2.50 \%}\\
    \hline
    \textcolor{black}{SSLGAN \cite{han2018semisupervised}} & \textcolor{black}{95.53 \%} & \textcolor{black}{90.35 \%} & \textcolor{black}{2.28 \%} & \textcolor{black}{4.76 \%}\\
    \hline
    \textcolor{black}{RBNet \cite{chen2017rbnet}} & \textcolor{black}{94.97 \%} & \textcolor{black}{91.49 \%} & \textcolor{black}{2.79 \%} & \textcolor{black}{4.99 \%}\\
    \hline
    \textcolor{black}{StixelNet-II \cite{garnett2017real}} & \textcolor{black}{94.88 \%} & \textcolor{black}{87.75 \%} & \textcolor{black}{4.04 \%} & \textcolor{black}{3.13 \%}\\
    \hline
    \textcolor{black}{RoadNet-RT} & \textcolor{black}{92.55 \%} & \textcolor{black}{93.21 \%} & \textcolor{black}{3.86 \%} & \textcolor{black}{7.84 \%}\\
    \hline
    \end{tabular}
\end{table}

\begin{figure*}[htbp]
    \centering
    \begin{subfigure}{0.33\textwidth}
            \includegraphics[width=\linewidth]{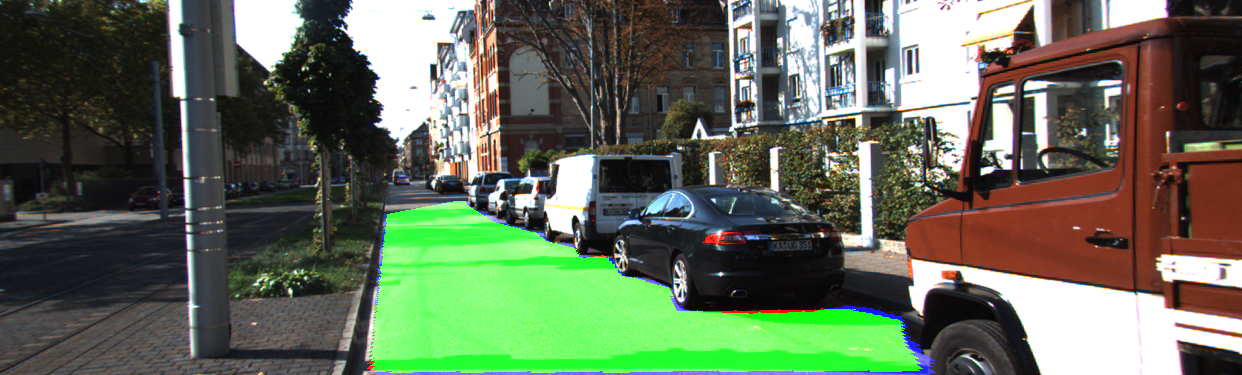}
    \end{subfigure}%
    \begin{subfigure}{0.33\textwidth}
            \includegraphics[width=\linewidth]{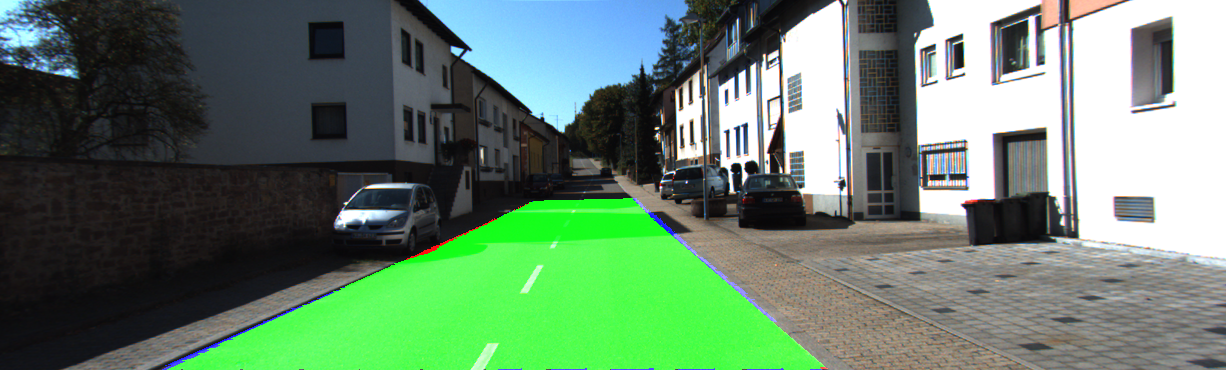}
    \end{subfigure}%
    \begin{subfigure}{0.33\textwidth}
            \includegraphics[width=\linewidth]{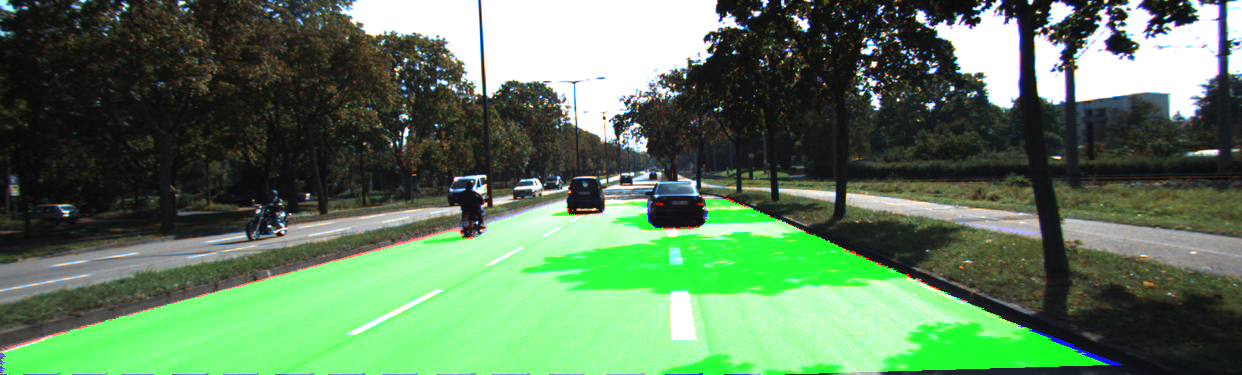}
    \end{subfigure}%
    \newline
    \begin{subfigure}{0.33\textwidth}
            \includegraphics[width=\linewidth]{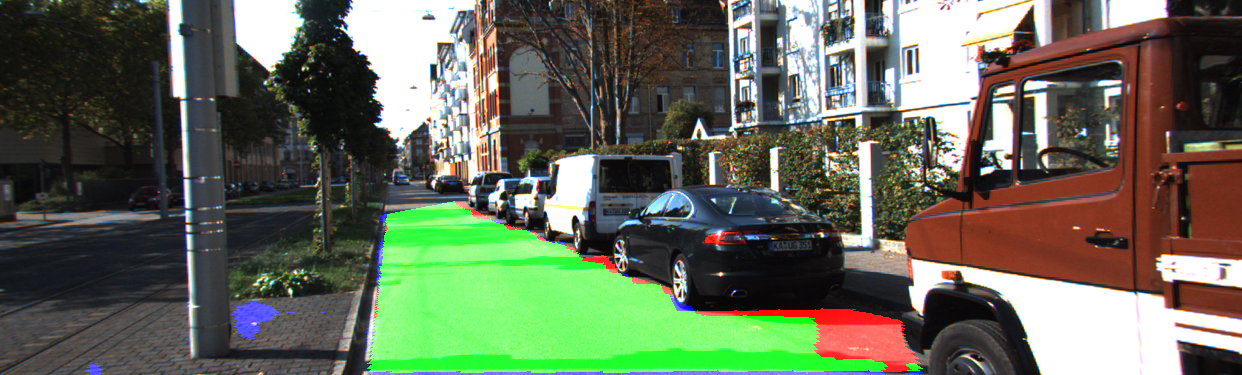}
    \end{subfigure}%
    \begin{subfigure}{0.33\textwidth}
            \includegraphics[width=\linewidth]{fig/mr_Persp_um_road_000077.png}
    \end{subfigure}%
    \begin{subfigure}{0.33\textwidth}
            \includegraphics[width=\linewidth]{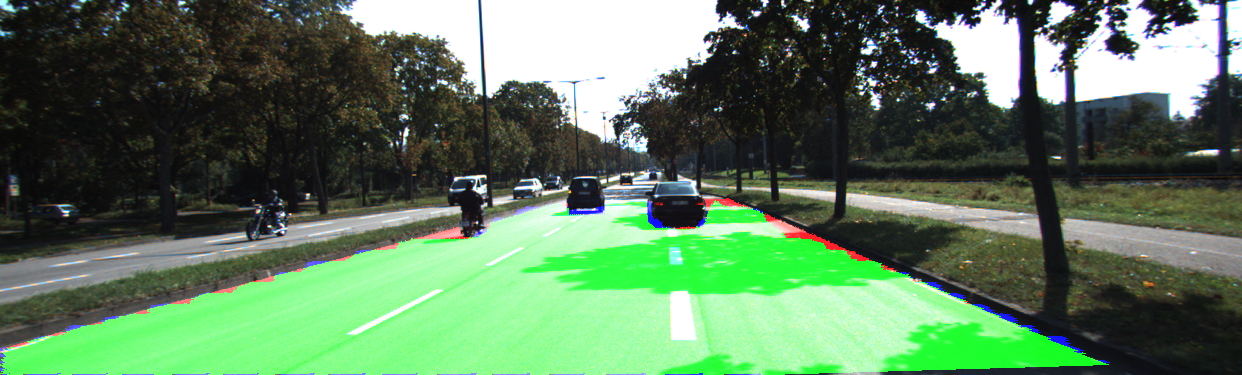}
    \end{subfigure}%
    \caption{\textcolor{black}{Comparison between RBANet (top) and RoadNet-RT (bottom)}}
    \label{fig:perf_comp}
\end{figure*}

\textcolor{black}{As shown in Fig.~\ref{fig:perf_comp}, comparing to RBANet, most of the classification errors of RoadNet-RT occurred near the boundary of the road since we choose not to include boundary attention in the model owing to the computation complexity. These errors won't affect autonomous driving due to path planning algorithm does not consider boundary of the drivable area.}

\textcolor{black}{\textbf{CamVid.}} \textcolor{black}{Besides the well-known KITTI dataset, the RoadNet-RT has also been evaluated on the CamVid dataset to verify its effectiveness on various road scenes (Tab.~\ref{tab:camvid_cmp}). For F1 score, RoadNet-RT achieves 92.98\% accuracy on CamVid test dataset, which is \textcolor{black}{3.74\%} less when comparing to the SOTA network RBANet \cite{sun2019reverse}. Since the processing time of RBANet on CamVid is not provided in \cite{sun2019reverse}, we skip the processing speed comparison on CamVid.}

\begin{table}[htbp]
\centering
    \caption{\textcolor{black}{Road segmentation results on the CamVid test dataset}}
    \begin{tabular}{|c|c|c|c|} 
        \hline
        \textcolor{black}{\textbf{Methods}} & \textcolor{black}{\textbf{F1-measure}} & \textcolor{black}{\textbf{Precision}} & \textcolor{black}{\textbf{Recall}}\\
        \hline
        \textcolor{black}{RBANet \cite{sun2019reverse}} & \textcolor{black}{96.72\%} & \textcolor{black}{97.14\%} & \textcolor{black}{96.30\%}\\
        \hline
        \textcolor{black}{RoadNet-RT} & \textcolor{black}{92.98\%} & \textcolor{black}{94.70\%} & \textcolor{black}{91.91\%}\\
        \hline
    \end{tabular}
\label{tab:camvid_cmp}
\end{table}

\begin{table*}[htbp]
    \centering
    \caption{\textcolor{black}{Performance evaluation from KITTI online test server}}
    \label{tab:perform}
    \begin{tabular}{c|c|c|c|c|c|c}
        \hline
        \textcolor{black}{\bf Benchmark} & \textcolor{black}{\bf MaxF} & \textcolor{black}{\bf AP} & \textcolor{black}{\bf PRE} & \textcolor{black}{\bf REC} & \textcolor{black}{\bf FPR} & \textcolor{black}{\bf FNR}\\
        \hline
        \textcolor{black}{UM\_ROAD} & \textcolor{black}{91.99\%} & \textcolor{black}{92.54\%} & \textcolor{black}{92.75\%} & \textcolor{black}{91.24\%} & \textcolor{black}{3.25\%} & \textcolor{black}{8.76\%}\\
        \hline
        \textcolor{black}{UMM\_ROAD} & \textcolor{black}{93.98\%} & \textcolor{black}{95.19\%} & \textcolor{black}{94.47\%} & \textcolor{black}{93.49\%} & \textcolor{black}{6.01\%} & \textcolor{black}{6.51\%}\\
        \hline
        \textcolor{black}{UU\_ROAD} & \textcolor{black}{90.79\%} & \textcolor{black}{91.67\%} & \textcolor{black}{91.79\%} & \textcolor{black}{89.80\%} & \textcolor{black}{2.62\%} & \textcolor{black}{10.20\%}\\
        \hline
        \textcolor{black}{URBAN\_ROAD} & \textcolor{black}{92.55\%} & \textcolor{black}{93.21\%} & \textcolor{black}{92.94\%} & \textcolor{black}{92.16\%} & \textcolor{black}{3.86\%} & \textcolor{black}{7.84\%}\\
        \hline
    \end{tabular}
\end{table*}
        

\section{Network Optimization for Hardware}
In this section, we summarize some guidelines to optimize specific CNNs toward FPGAs accelerator implementation. So that on-chip resources efficiency and computation efficiency FPGA design are maximized. Different from the conventional optimization techniques, the goal of this step is to balance the number of operations, number of weights and computation patterns, while remaining the accuracy within a reasonable range.

\subsection{Depthwise Separable Convolution}
\textcolor{black}{Depthwise separable convolution is initially introduced in \cite{sifre2014rigid}. It has been widely adopted by a great number of lightweight neural networks such as Xception \cite{chollet2017xception}, MobileNet series \cite{howard2017mobilenets}\cite{sandler2018mobilenetv2}. The main idea of depthwise separable convolution is to decompose standard convolution into a 3$\times$3 depthwise convolution and a 1$\times$1 pointwise convolution to achieve smaller number of weights and consequently less operations. Assuming $D_K$ is the size of convolution kernel, $M$ is the depth of input feature maps and $N$ is the number of convolution kernels (also the channel number of output feature maps).}

During depthwise convolution, a single filter is applied to each input channel. And then the pointwise convolution applies a 1$\times$1 convolution to combine the outputs of the depthwise convolution. The number of weights required by standard convolution and depthwise separable convolution are calculated in (\ref{eq:std_conv}) and (\ref{eq:dw_conv}) respectively.
\begin{equation}
    D_K\! \cdot\! D_K\cdot\! M\! \cdot\! N
    \label{eq:std_conv}
\end{equation}
\begin{equation}
    D_K\! \cdot\! D_K\!\cdot\! M +\! M\!\cdot\! N
    \label{eq:dw_conv}
\end{equation}
Therefore, when replacing standard convolution with depthwise separable convolution, the reduction ratio of weights is 
\begin{equation}
    \frac{D_K\! \cdot\! D_K\!\cdot\! M+\! M\!\cdot\! N}{D_K\! \cdot\! D_K\!\cdot\! M\! \cdot\! N}=\frac{1}{N}+\frac{1}{D_K^2}
    \label{eq:pw_conv}
\end{equation}

Besides the parameter reduction and operation number decreasing, from the hardware implementation point of view, 
depthwise separable convolution need not as large size accumulator as required by standard convolution. In standard convolution, every element of output feature map is the sum of $D_K\!\cdot D_K\!\cdot M$ elements. While in depthwise separable convolution, that is the sum of $D_K\!\cdot D_K$ and $M$ elements for depthwise convolution and pointwise convolution respectively.
\textcolor{black}{On the other side, separating standard convolution into depthwise convolution and point convolution requires intermediate feature map buffering, and hence demands larger bandwidth.}

\textcolor{black}{Applying this to RoadNet-RT proposed in this paper, the total number of parameters is reduced from 756K to 134K}, which is illustrated in Tab.~\ref{tab:dsep_conv}. Although the accuracy loss is 1.37\%, the number of parameters reduces by a factor of 5.64.

\begin{table}[htbp]
    \centering
    \caption{\textcolor{black}{Comparison of RoadNet-RT with and without depthwise separable convolution}}
    \label{tab:dsep_conv}
    \begin{tabular}{c|c|c}
        \hline
         Convolution type & IOU\footnotemark[1] & parameters\\
        \hline
         Standard & \textcolor{black}{93.67$\%$} & 756,032\\
        \hline
         Depthwise separable & \textcolor{black}{92.30$\%$} & 133,870\\
        \hline
    \end{tabular}
\end{table}
\footnotetext[1]{Since KITTI online test sever limits the submission to be 3 times per month, therefore 20\% of the training set has been split as validation set to evaluate the methods we proposed. Here we choose IOU as the main metric to estimate the performance of different methods. IOU is one of the most important and the most widely used metrics for segmentation performance evaluation.}

\subsection{Large kernel size convolution}
The most commonly used kernel size for convolution is 3$\times$3. However, in order to have large size of field of perception, especially in the first layer, large kernel size is usually desired (7$\times$7 in ResNet \cite{he2016deep} for instance).


\begin{algorithm}
    \caption{Cascaded loop of standard convolution}
    \label{alg:cas_loop}
\begin{algorithmic}[htbp]
    \For{no in Nof}\Comment{output channel,loop-4}
        \For{(y,x) in (Noy,Nox)}\Comment{feature map,loop-3}
            \For{ni in Nif}\Comment{input channel,loop-2}
                \For{(ky,kx) in (K,K)}\Comment{kernel,loop-1}
                    \State $F_{out}$[no,y,x]+=
                    \State $F_{in}$[ni,y-ky,x-kx] *$K$[no,ni,ky,kx]
                \EndFor
            \EndFor
        $F_{out}$ += $bias$[no]
        \EndFor
    \EndFor
\end{algorithmic}
\end{algorithm}

However, to deal with different kernel size filters affects either parallelism of processing or the efficiency of buffer usage. From matrix multiplication point of view (in Alg.~\ref{alg:cas_loop}), through keeping the loop-1, hardware accelerator can handle different size of filters without extra multipliers consumed. But the penalty is the parallelism of loop-1 loss. However, different size of filter requires different size of on-chip memory. Consider a feature map with size $W\!\cdot H\!\cdot C$, to buffer it for $K\!\cdot K$ filter, memory size $(W\! +\!K\!-\!1)\!\cdot (H+K-1)\!\cdot C$ is need. So that the feature map buffer for 7$\times$7 filter is $4\cdot(W\! +\! H\! +\! 4)/(W\!\cdot H)$ times larger than that for 3$\times$3 filter.

\begin{figure}[htbp]
    \centering
    \includegraphics[width=0.95\columnwidth]{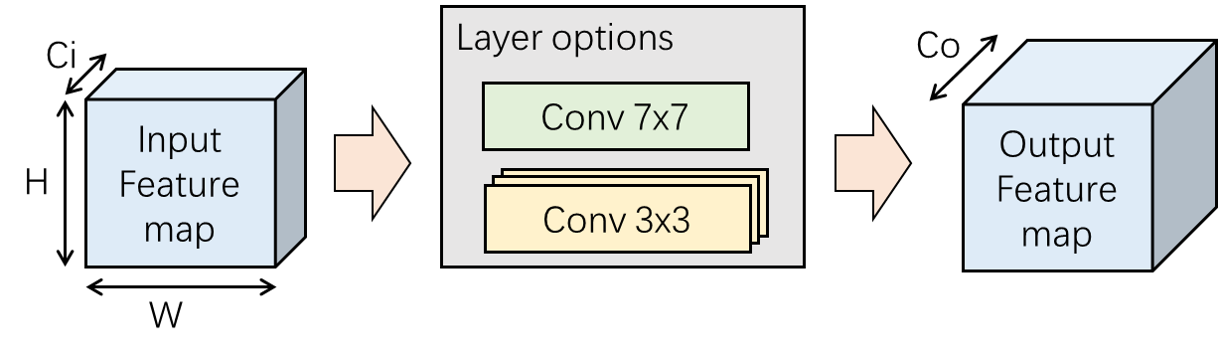}
    \caption{Strategy for large convolutional layer replacement}
    \label{fig:largeconv_replace}
\end{figure}

To pursue the same perceptive field of $7\times 7$, three cascaded convolutional layers with kernel size $3\times 3$ can replace one convolutional layer with kernel size $7\times 7$. If so, there is no extra resource needed including both multipliers and memory. Besides, the number of operations decreases. As illustrated in Fig.~\ref{fig:largeconv_replace}, for input feature map size $W\!\cdot H\!\cdot C_i$ and output feature map size $W\!\cdot H\!\cdot C_o$, if 7$\times$7 filter is applied, totally $(W\!\cdot H\!\cdot 7\!\cdot 7\times C_i\!\cdot C_o)=49\!\cdot W\!\cdot H\!\cdot C_i\!\cdot C_o$ GOPS costs. In case of three $3\times 3$ convolutional layers, $3\!\cdot (W\!\cdot H\!\cdot 3\!\cdot 3\!\cdot C_i\!\cdot C_o))=27\!\cdot W\!\cdot H\!\cdot C_i\!\cdot C_o$.

The performance comparison between these two options mentioned above is shown in Tab.~\ref{tab:large_conv_cmp}. When replacing the first convolutional layer (7$\times$7) with three 3$\times$3 convolutional layers, the accuracy loss in IOU is 0.19\%. Since there is only one layer of 7$\times$7 convolution, the save in operations and parameters are negligible.

\begin{table}[htbp]
    \centering
    \caption{\textcolor{black}{Comparison between 7$\times$7 convolution and its replacement ($C_i$ is the input feature map channel number and $C_o$ is the output feature map channel number, they equal 32 and 64 respectively in this experiment)}}
    \label{tab:large_conv_cmp}
    \begin{tabular}{c|c|c}
        \hline
         Method & IOU & parameter \\
        \hline
         1 conv $7\times 7$ & \textcolor{black}{93.67$\%$} & $7\!\cdot\! 7\!\cdot\! C_i\!\cdot\! C_o$ \\
        \hline
         3 conv $3\times 3$ & \textcolor{black}{93.67$\%$} & $3\!\cdot\!3\!\cdot\!C_i\!\cdot\! C_o\!+\!3\cdot\! 3\!\cdot\! C_o\!\cdot\! C_o+3\!\cdot\! 3\!\cdot\! C_o\!\cdot\! C_o$\\
        \hline
    \end{tabular}
\end{table}

In the segmentation networks, dilated convolution \cite{yu2015multi} is the most widely used method to enlarge the perceptive field without introducing more weights. Unfortunately, during convolution with dilated kernel (3$\times$3 with dilated rate equals 3 for instance), the region required from feature map is still 7$\times$7. This will introduce the dilemma described above still. The only difference is, if using three 3$\times$3 convolutional layers instead of one dilated 3$\times$3 convolutional layers with dilated rate as 3, two times more weights and two times more operations are unavoidable. However, since the dilated convolutional layer usually won't dominant, this penalty is still affordable.

\begin{table}[htbp]
    \centering
    \caption{\textcolor{black}{Performance comparison between dilated convolution ($3\times 3$ with dilated rate 3) and its replacement}}
    \label{tab:dilat_conv_cmp}
    \begin{tabular}{c|c|c}
        \hline
         Method & IOU & parameter \\
        \hline
        1 conv $3\times 3$ & \textcolor{black}{\multirow{2}{3em}{93.67$\%$}} & \multirow{2}{4em}{$3\!\cdot\! 3\!\cdot\! C_i\!\cdot\! C_o$}\\
        dilated rate 3 & & \\
        \hline
         3 conv $3\times 3$ & \textcolor{black}{93.66$\%$} & $3\!\cdot\!3\!\cdot\!C_i\!\cdot\! C_o\!+\!3\cdot\! 3\!\cdot\! C_o\!\cdot\! C_o+3\!\cdot\! 3\!\cdot\! C_o\!\cdot\! C_o$ \\
        \hline
    \end{tabular}
\end{table}

\subsection{Consideration of channel depth}
In our hardware implementation, after considering the given resources on ZCU102 board, loop-2 in Alg.~\ref{alg:cas_loop} has been unrolled with 32 feature maps processed in parallel. To maximum the computation efficiency of accelerator, it's better that the input feature map depth of all layers align to integer factor of 32.

\subsection{Batch Normalization}
During inference, Batch Normalization (BN) is downgraded into 1$\times$1 convolution and further merged into convolutional layer prior than it. The merged weights and bias follow (\ref{eq:w}) and (\ref{eq:b}), where $W_{}$ and $b_{}$ represent weights and bias respectively.

\begin{align}
    W_{merge} = W_{B\!N}\cdot W_{conv} \label{eq:w}\\
    W_{merge} = W_{B\!N}\cdot b_{conv}+b_{B\!N}\label{eq:b}
\end{align}

Batch normalization layer is helpful for fast convergence but not always a necessary layer concerning to the accuracy (PointNet\cite{qi2017pointnet} for instance). The contribution of BN layer is evaluated in Tab.\ref{tab:opt_bn}, from which we find in our segmentation neural network, BN helps to increase the accuracy by \textcolor{black}{0.28$\%$} without too much difference in convergence. Therefore, BN layers are kept in RoadNet-RT.

\begin{table}[htbp]
    \centering
    \caption{\textcolor{black}{The performance comparison with and without BN layer, both of them are trained using the same batch size and the sam GPU}}
    \label{tab:opt_bn}
    \begin{tabular}{c|c|c}
        \hline
         Method & with BN & without BN\\
        \hline
         IOU & \textcolor{black}{93.67$\%$} & \textcolor{black}{93.39$\%$} \\
        \hline
         converge@epoch & ~350 & ~340\\
        \hline
         duration/epoch & ~10s & ~8s\\
        \hline
    \end{tabular}
\end{table}

Some experiments declared that BN after ReLU usually shows better result \cite{bnrelu}. But this may vary from one network to another.

\subsection{Quantization}
To maximize the computation capability of FPGA, fixed point operations is preferred. Quantization aware training has been performed for 8-bit and 16-bit respectively with the help of model optimization library from QKeras \cite{QKeras}. Brute-force quantization may lead to unacceptable precision loss. While quantization aware training restricts the bit-width during training. This not only compensates the precision loss but introduces more non-linearity.

\textcolor{black}{The performance after quantization is shown in Tab.~\ref{tab:quant}. The IoU accuracy of 8-bit implementation is 92.36$\%$, while that of 16-bit quantization is 92.40$\%$. The accuracy of 16-bit quantization is 0.04$\%$ higher than that of 8-bit quantization, but it requires twice much memory for weights storage. Here we choose the 8-bit INT quantization for hardware implementation}, 1) from storage perspective, memory space for 8-bit weights is only half of that for 16-bit quantization, 2) from hardware resources perspective, each DSP48E2 core could perform two 8-bit multiplications simultaneously but only one for 16-bit multiplication \cite{fu2017deep}.

\begin{table}[htbp]
    \centering
    \caption{\textcolor{black}{Performance of 8-bit and 16-bit quantized networks}}
    \label{tab:quant}
    \begin{tabular}{c|c|c}
        \hline
         Bit Width & IOU & size of parameters\\
        \hline
         float32 & 93.67\% & 2.88MB\\
        \hline
         int16 & 92.40\% & 1.44MB\\
        \hline
         int8 & 92.36\% & 0.72MB\\
        \hline
    \end{tabular}
\end{table}

\subsection{\textcolor{black}{Progressive impact of optimization techniques}}
\textcolor{black}{Considering the impact on precision loss, all the optimization (opt) techniques described above have been applied to RoadNet-RT progressively. The corresponding changes in IOU precision are listed in Tab.~\ref{tab:ablation}. As mentioned earlier, 7$\times$7 kernel can be computed using 3$\times$3 convolutions and there is no degradation of accuracy. Next, we use 3$\times$3 convolution to replace dilated convolution with different dilated rates. There are four dilated convolutions in RoadNet-RT, which accounts for a performance drop to 0.05\%. Depthwise separable convolution sharply compresses the computation complexity at the penalty of reduced network capacity, resulting an additional (0.55\%) precision loss. Finally we apply fixed-point quantization to the model, which contributes to the largest precision loss (1.08\%) among all optimization techniques.}

\begin{table}[htbp]
    \centering
    \caption{\textcolor{black}{Performance comparison for different techniques}}
    \label{tab:ablation}
    \begin{tabular}{l|l}
        \hline
         \textcolor{black}{Technique} & \textcolor{black}{IOU} \\
        \hline
         \textcolor{black}{No optimization} & \textcolor{black}{93.67$\%$} \\
        \hline
         \textcolor{black}{opt 1 - Replace 7$\times$7 kernel } & \textcolor{black}{93.67$\%$} \\
        \hline
         \textcolor{black}{opt 2 - Replace dilated convolution } & \textcolor{black}{93.62$\%$} \\
        \hline
         \textcolor{black}{opt 3 - Depthwise separable convolution} & \textcolor{black}{93.07$\%$} \\
        \hline
         \textcolor{black}{opt 4 - Quantization (INT8 on FPGA)} & \textcolor{black}{91.99$\%$} \\
        \hline
    \end{tabular}
\end{table}

\section{System-on-Chip Implementation}
To fully utilize the computation resources, the whole system is partitioned into software part (done by ARM processor) and hardware part (running on FPGA). The software part job is image resize for both input and output of neural network (Fig.~\ref{fig:roadnet}). With the help of OpenCV library \cite{2015opencv}, image resize can be easily done on PYNQ platform. 

\begin{figure}[htbp]
    \centering
    \includegraphics[width=0.75\columnwidth]{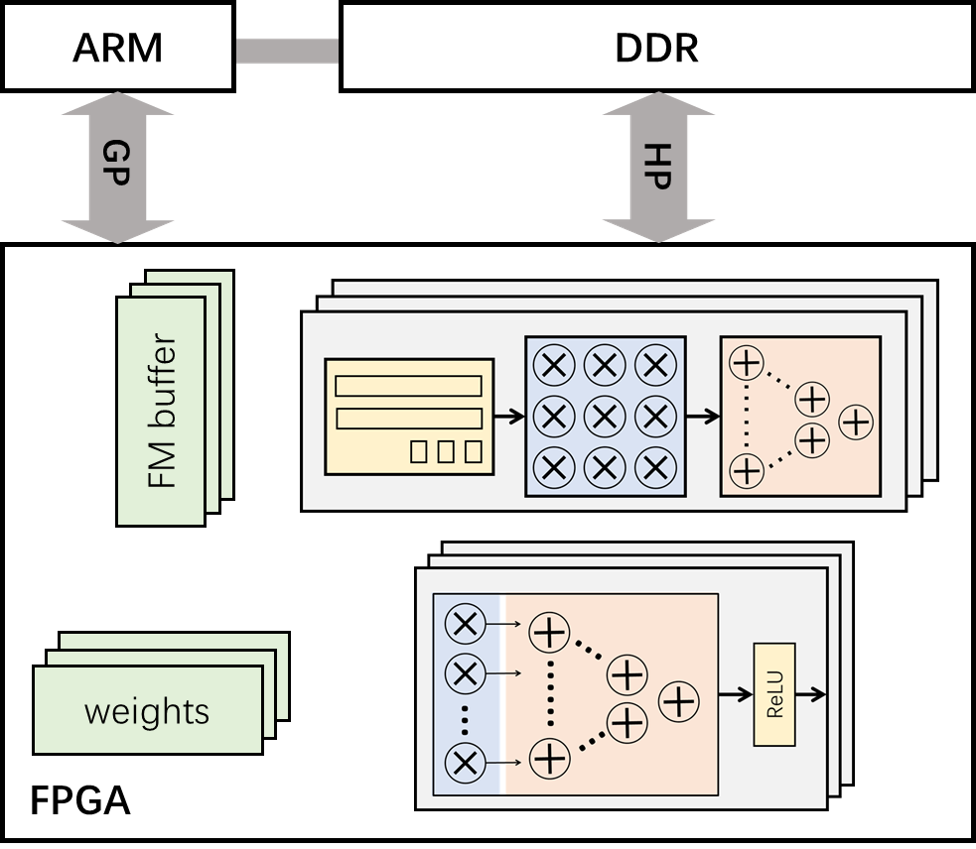}
    \caption{System overview of RoadNet-RT accelerator}
    \label{fig:hw_arch}
\end{figure}

\textcolor{black}{Combining the depthwise and pointwise convolution into one process engine array is possible, but may result in output feature map reshape before sending to DDR memory and consequently decrease the efficiency of the accelerator. Thereby, we decide to separately implement these two computation modules.} The overview of hardware architecture is demonstrated in Fig.~\ref{fig:hw_arch}. It consists of depthwise convolution module, and pointwise convolution module, feature map buffers, weights buffers. A finite state machine controls the running order of CNN operations. All the modules mentioned above are configurable based on the on-chip resources available on the target FPGA platform.
 
\textcolor{black}{We chose 32 as the depth of process engine array, due to 1) target ZCU102 development kit supplies 2520 DSPs and 32.1Mb BRAM, which is sufficient for 32 process engines and corresponding feature map buffers 2) considering except the input layer and output layer, the depth of all the layers in RoadNet-RT are the product of 32, therefore using 32 can maximize the utilization of each multiplier, 3) as the greatest common divisor, using 32 as depth can minimize the data transporting for convolutions whose depth is large.}

\subsection{Depthwise convolution module}
\begin{figure}[htbp]
    \centering
    \includegraphics[width=0.75\columnwidth]{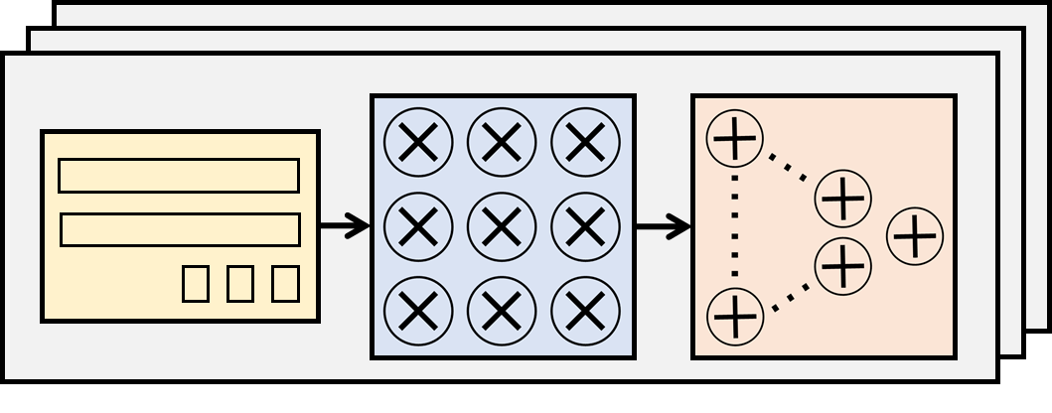}
    \caption{Block diagram of depthwise convolution module}
    \label{fig:depth_conv_arch}
\end{figure}

Depthwise convolution module (Fig.~\ref{fig:depth_conv_arch}) contains line buffers, process engines (PEs) and adder trees. As descried in the previous section, to unroll the kernel loop (loop-1 in Alg.~\ref{alg:cas_loop}), line buffer is needed to generate the sliding patch. Since kernel size of all the convolutional layers in this segmentation network is $3\times 3$, a multiplier array with length equal to 9 follows the line buffer. Correspondingly, an adder tree in the end sums the products up. To balance the computation efficiency and on-chip resources, the batch size of depthwise convolution module is set to 32.

\subsection{Pointwise convolution module}
\begin{figure}[htbp]
    \centering
    \includegraphics[width=0.75\columnwidth]{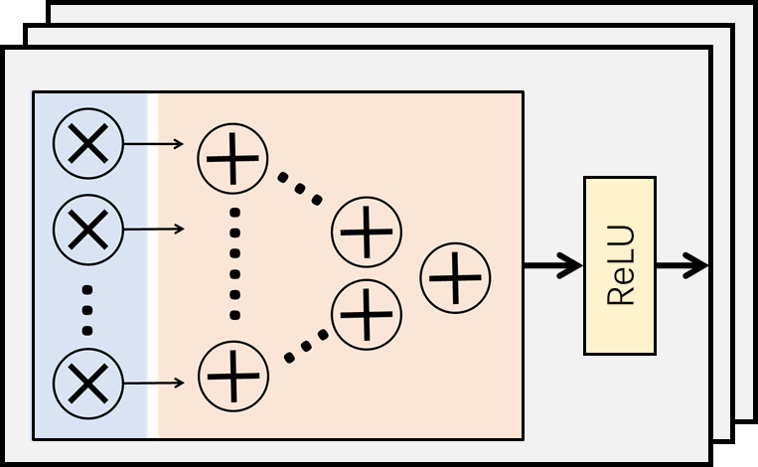}
    \caption{Block diagram of pointwise convolution module}
    \label{fig:point_conv_arch}
\end{figure}

To align to the depthwise convolution module to fit the same size of feature buffers, the pointwise convolution module (Fig.~\ref{fig:point_conv_arch}) is designed to handle $32\times 1$ vector - $32\times 32$ matrix multiplication. There are 3 components multiplier array, adder tree, and ReLU module form the Pointwise convolution module. If the batch normalization layer is placed before ReLU layer, it can be merged and completed by multiplier array and adder tree. Otherwise, 1 extra multiplier and 1 extra adder is necessary to perform the batch normalization operation.

\subsection{GAM Module and FFM Module}
Both GAM and FFM modules require operations with totally different computation patterns. Global average pooling is to calculate the average value of one entire channel. Therefore, an accumulator plus one multiplier for each channel has been implemented. The following $1\times 1$ convolution is mathematically vector-matrix multiplication, which can be either routed into pointwise convolutional module or implemented with extra resource, given the resource consumption of this operation is small. Sigmoid function is approximated by the piece-wise function and implemented using a Look-Up Table.

\subsection{Buffers}
The on-chip memory are divided into buffers for feature maps, weights and global pooling result respectively. In this design, 1) there is no biases, so that no extra buffer is needed for bias storage, and 2) since the weights occupy only small portion of the on-chip memory, so that they can be hard coded into on-chip memory.

To boost the processing speed, one effective way is to reduce the number of time data transmission (between FPGA and DDR memory). Multiple feature map buffers with size \textcolor{black}{35$\times$120$\times$32} have been implemented as ping-pong buffers to decrease data swap as much as possible.

\subsection{Tasks on ARM Processor}
\begin{figure}[htbp]
    \centering
    \includegraphics[width=0.8\columnwidth]{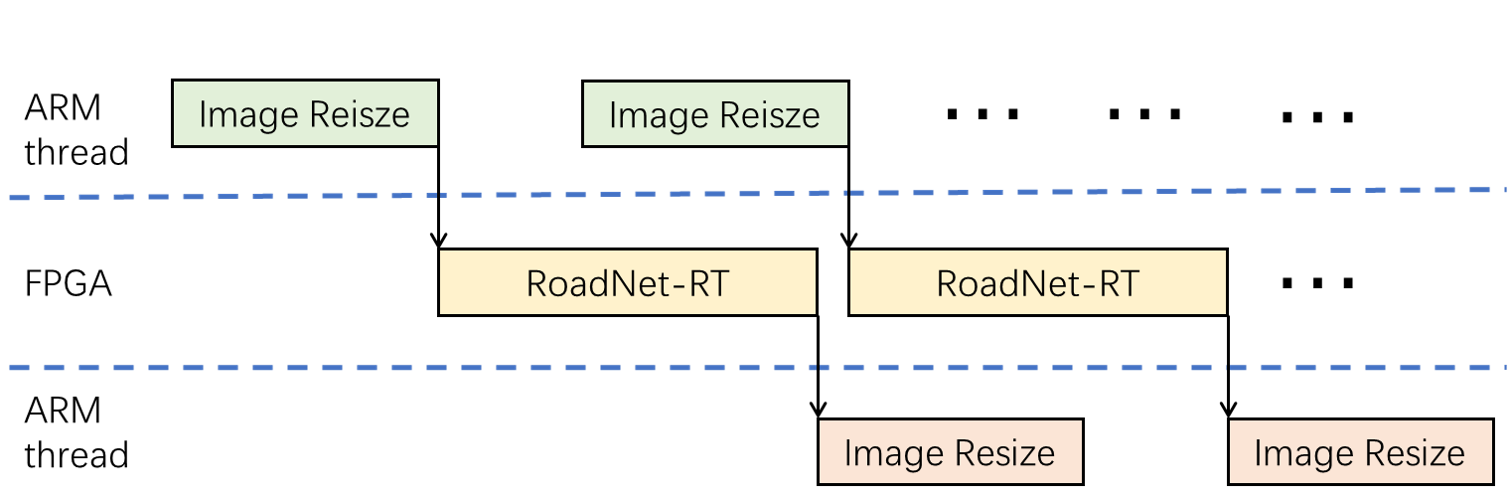}
    \caption{Task partitioning of RoadNet-RT on SoC}
    \label{fig:partition}
\end{figure}

\textcolor{black}{Referring to Fig.~\ref{fig:hw_arch}, the entire CNN is implemented on FPGA side. In order to fully utilize the available computation resources on SoC, the rest of the task has been assigned to the ARM processor. Thus, the whole RoadNet-RT are partitioned to both ARM processor and FPGA as shown in Fig.~\ref{fig:partition}. All the three tasks are overlapped and pipelined, and this consequently speeds up the system speed.}

\section{Results and Discussion}
\begin{figure}[htbp]
    \centering
    \includegraphics[width=0.8\columnwidth]{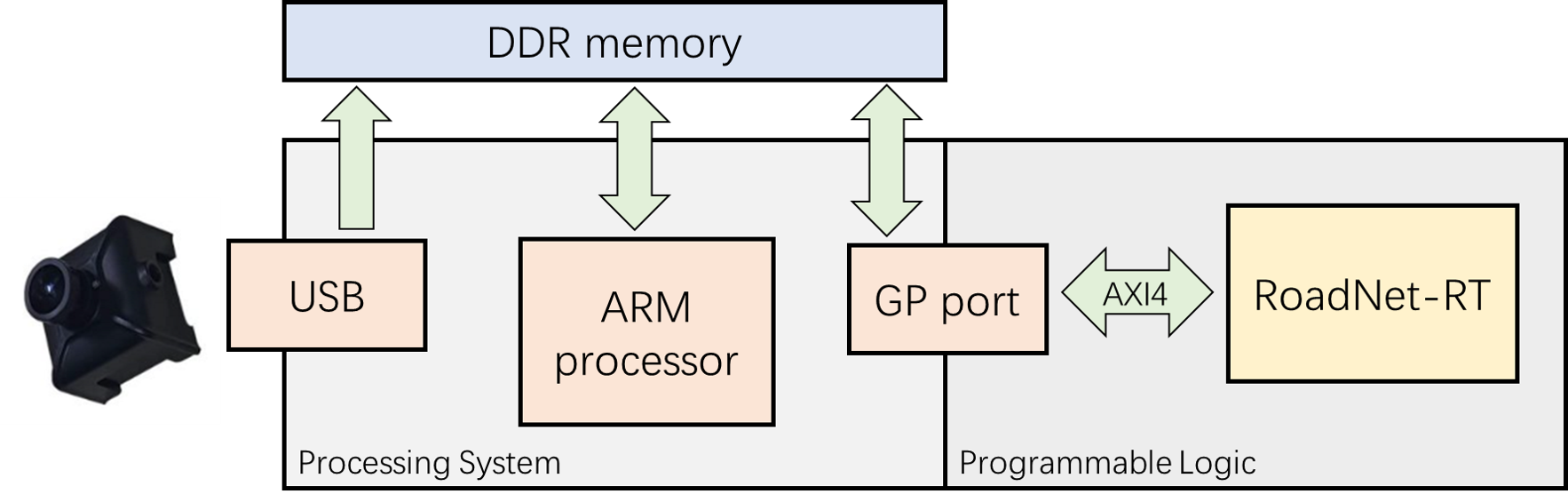}
    \caption{Setup of the road segmentation system}
    \label{fig:setup}
\end{figure}

The implementation tools used in this paper are Xilinx Vivado HLS and MATLAB HDL Coder Toolbox. The whole system has been implemented on ZCU102 development kit, with the PYNQ system installed (The system setup is show in Fig.~\ref{fig:setup}). There are 548,160 Flip-Flops (FFs), 274,080 Look-Up Tables (LUTs), 1824 (32.1 Mb) Block RAMs (BRAMs) and 2,520 DSPs on the board. The FPGA resources consumption of this accelerator for both 16-bit and 8-bit quantization formats are shown in Tab.~\ref{tab:resource}.

\begin{table}[htbp]
    \centering
    \caption{\textcolor{black}{FPGA on-chip resource usage of RoadNet-RT}}
    \label{tab:resource}
    \begin{tabular}{c|c|c|c|c}
        \hline
         bitwidth & FF & LUT & DSP & BRAM\\
        \hline
         8-bit & 115684 & 260335 & 1560 & 1340\\
        \hline
    \end{tabular}
\end{table}

Since each DSP48E2 slice can handle two $8$-bit$\times8$-bit multiplication while the number for $16$-bit number is one, thus $8$-bit format accelerator consumes almost the same DSP slices and BRAMs as that in $16$-bit format but twice the number of input images. To maximum the computation capability of hardware, we quantize all the weights into 8-bit. \textcolor{black}{When running at 250 MHz, this 8-bit accelerator's processing speed is 196.7 fps.} In Tab.~\ref{tab:road_compare}, all the image-based road segmentation solutions in the KITTI leaderboard are summarized and compared to our solution in GPU and FPGA. Most of the existing methods cost $100$ ms or longer. One of the only two real-time solutions FCN-LC \cite{mendes2016exploiting} runs on TITAN X GPU, which requires 600-650W power supply on PC to support. Therefore, our solutions supply a well-balanced and practical way to run this the road segmentation task on embedded devices.

In this accelerator, there are \textcolor{black}{8} feature map buffers are allocated. But this number may vary according to the balance between available resources on the target FPGA and required processing speed. More feature map buffers can store more intermediate feature maps and consequently increase the processing speed. While less feature map buffers require more temporary data stored in external memory rather than on-chip ones. And thus leads to longer processing time.

\begin{table}[htbp]
    \centering
    \caption{\textcolor{black}{Performance comparison between floating point and fixed point}}
    \label{tab:fpga_perf}
    \begin{tabular}{c|c|c|c}
        \hline
         Device & precision & accuracy (IOU) & processing time\\
        \hline
         GPU & float32 & 93.67\% & 9 ms\\
        \hline
         FPGA & int8 & \textcolor{black}{91.99\%} & 5.24 ms\\
        \hline
    \end{tabular}
\end{table}

\begin{table*}[htbp]
	\caption{Performance comparison of all the image-based road segmentation solutions in the KITTI leaderboard (blank means it is not mentioned in the original paper)}
	\label{tab:road_compare}
	\centering
	\begin{tabular}{|c|c|c|c|c|c|}
	    \hline
	    Name & CNN-based & Input shape & Devices & Accuracy(MaxF) & Processing speed \\
	    \hline
	    RBANet\cite{sun2019reverse} & \ding{51} & $360\times 720$ & TITAN XP & 96.30\%  & 160 ms\\
	    \hline
	    SSLGAN\cite{han2018semisupervised} & \ding{51} & $375\times 1242$ & TITAN X & 95.53\% & 700 ms\\
	    \hline
	    RBNet\cite{chen2017rbnet} & \ding{51} & $300\times 900$ & Tesla K20c & 94.97\% & 180 ms\\
	    \hline
	    StixelNet-II\cite{garnett2017real} & \ding{51} & $800\times 370$ & Quadro M6000 & 94.88\% & 1200 ms\\
	    \hline
	    MultiNet\cite{teichmann2018multinet} & \ding{51} & $1248\times 384$ &  & 94.88\% & 170 ms\\
	    \hline
	    RoadNet3\cite{lyu2019road} & \ding{51} & $600\times 160\times 5$ & GTX 950M & 94.44\% & 300 ms\\
	    \hline
	    DEEP-DIG\cite{munoz2017deep} & \ding{51} &  & Titan X & 93.98\% & 140 ms\\
	    \hline
	    Up-Conv-Poly\cite{leivas2016ef} & \ding{51} & $500\times 500$ & TITAN X & 93.83\% & 83 ms\\
	    \hline
	    OFA-Net\cite{zhang2020one} & \ding{51} &  &  & 93.74\% & 40 ms\\
	    \hline
	    Up-Conv\cite{leivas2016ef} & \ding{51} & $300\times 300$ & GTX TITAN X & 92.39\% & 52.2 ms\\
	    \hline
	    ALO-AVG-MM\cite{reis2019combining} & \ding{51} & $624\times 192$ & GTX 1080  & 92.03\% & 29.6 ms\\
	    \hline
	    FTP\cite{laddha2016map} & \ding{51} & & & 91.61\% & 280 ms\\
	    \hline
	    PT-ResNet\cite{fan2019pt} & \ding{51} &  & GTX 1080 Ti & 91.61\%  & 300 ms\\
	    \hline
	    FCN-LC\cite{mendes2016exploiting} & \ding{51} & $621\times 187$ & TITAN X & 90.79\% & 30 ms\\
	    \hline
	    StixelNet\cite{levi2015stixelnet} & \ding{51} & $24\times 370$ &  & 89.12\% & 1000 ms\\
	    \hline
	    MAP\cite{laddha2016map} & \ding{51} &  &  & 87.80\%  & 280 ms\\
	    \hline
	    SPRAY\cite{kuhnl2012spatial} & \ding{51} & $800\times 600$ & GTX 580 &  87.09\% & 45 ms\\
	    \hline
	    multi-task CNN\cite{oeljeklaus2018fast} & \ding{51} & $375\times 1242$ & unknown type GPU & 86.81\%  & 25.1 ms\\
	    \hline
	    PGM-ARS\cite{passani2015fast} & \ding{51} & $\sim 75\times 248$ & Intel i7-4700MQ processor  & 85.69\% & 50 ms\\
	    \hline
	    SRF\cite{xiao2016monocular} & \ding{55} & $500\times 250$ &  & 82.44\% & 200 ms\\
	    \hline
	    ARSL-AMI\cite{passani2014crf} & \ding{55} &  &  & 80.36\%  & 50 ms\\
	    \hline
	    CN\cite{alvarez2012road} & \ding{55} &  &  & 79.02\% & 2000 ms\\
	    \hline
	    \textcolor{black}{\textbf{Ours}} & \textcolor{black}{\ding{51}} & \textcolor{black}{$\textbf{280}\times \textbf{960}$} & \textcolor{black}{\textbf{GTX 1080}} & \textcolor{black}{\textbf{92.55\%}} & \textcolor{black}{\textbf{9 ms}}\\
	    \hline
	\end{tabular}
\end{table*}

\textcolor{black}{The FPGA performance on the KITTI valid dataset is shown in Tab.~\ref{tab:fpga_perf}. After replacing all the large kernel, dilated convolution into convolutions with uniform kernel size and quantization, when using INT8 format weights, the IOU of network on FPGA is \textcolor{black}{91.99\%}, which is \textcolor{black}{1.68\%} less than the proposed floating point RoadNet-RT.}

\section{Conclusion}
This paper presents a real-time, high-throughput convolutional neural network architecture for road segmentation. Several optimization techniques are applied to reduce the number of operations while preserving the accuracy performance. \textcolor{black}{This networks achieves $92.55\%$ MaxF score on KITTI dataset with $111$ fps on GTX 1080 GPU (for image size $280\times 960$).} More importantly, using RoadNet-RT as an example, we present a systematic approach on how to perform CNN network optimization for hardware implementation. Following this as a guideline, one can easily convert any existing CNN structure into a computation efficient, high-throughput architecture for FPGA with little loss in accuracy. \textcolor{black}{Several experiments have been conducted to support the proposed approach. In the end, a SoC design has been successfully demonstrated on ZCU102 FPGA development kit, which speeds up the processing time by a factor of $1.72$ comparing to its GPU implementation.}


\ifCLASSOPTIONcaptionsoff
  \newpage
\fi



\bibliographystyle{IEEEtran}
\bibliography{roadSegFPGA}
%
%
%

%







\end{document}